\documentclass{emulateapj}

\usepackage{natbib}
\usepackage{lscape}
\usepackage{rotating}

\usepackage{hyperref}

\usepackage{amsmath}
\submitted{draft version \today}

\def\kpc{{\rm kpc}}
\def\ifm#1{\relax\ifmmode#1\else$\mathsurround=0pt #1$\fi}

\def\msun{M_{\odot}}

\def\ltsima{$\; \buildrel < \over \sim \;$}
\def\lsim{\lower.5ex\hbox{\ltsima}}
\def\gtsima{$\; \buildrel > \over \sim \;$}
\def\gsim{\lower.5ex\hbox{\gtsima}}

\def\ie{{\it i.e.}\,}
\newcommand{\kms}{\, {\rm km\, s}^{-1}}

\newcommand{\mypm}[2]{^{+#1}_{-#2}}

\shorttitle{SL2S: modeling of pilot SL2S galaxy-scale lens sample}
\shortauthors{Gavazzi et~al.}

\def\iap{1}
\def\ucsb{2}
\def\oxford{3}
\def\kipac{4}
\def\umel{5}

\def\Nlens{16}
\def\Nhst{10}

\begin{document}

\title{The SL2S Galaxy-scale Gravitational Lens Sample. I. The alignment of mass and light in massive early-type galaxies at $z=0.2-0.9$}

\author{Rapha\"el~Gavazzi\altaffilmark{\iap}}
\author{Tommaso~Treu\altaffilmark{\ucsb}$^{\dag}$}
\author{Philip~J.~Marshall\altaffilmark{\oxford,\kipac,\ucsb}}
\author{Florence~Brault\altaffilmark{\iap}}
\author{Andrea~Ruff\altaffilmark{\umel}}


\altaffiltext{\iap}{Institut d'Astrophysique de Paris, UMR7095 CNRS -- Universit\'e Pierre et Marie Curie, 98bis bd Arago, F-75014 Paris, France}
\altaffiltext{\ucsb}{Physics Department, University of California, Santa Barbara, CA 93106, USA} 
\altaffiltext{\oxford}{Department of Astrophysics, Oxford University, Denys Wilkinson Building, Keble Road, Oxford OX1 3RH, UK}
\altaffiltext{\kipac}{KIPAC, P.O. Box 20450, MS29, Stanford, CA 94309, USA}
\altaffiltext{\umel}{School of Physics, Astrophysics, University of Melbourne, Parkville 3010, Victoria, Australia}
\altaffiltext{$\dag$}{{Alfred P. Sloan Research Fellow}}


\begin{abstract}
We study the relative alignment of mass and light in a sample of
\Nlens\ massive early-type galaxies at $z=0.2-0.9$ that act as strong
gravitational lenses. The sample was identified from deep multi-band
images obtained as part of the Canada France Hawaii Telescope Legacy Survey
as part of the Strong Lensing Legacy Survey (SL2S). Higher
resolution follow-up imaging is available for a subset of \Nhst\
systems. We construct gravitational lens models and infer total
enclosed mass, elongation, and position angle of the mass
distribution. By comparison with the observed distribution of light we
infer that there is a substantial amount of external shear with
mean value $\langle\gamma_{\rm ext}\rangle= 0.12 \pm 0.05$, arising
most likely from the environment of the SL2S lenses. In a companion
paper, we combine these measurements with follow-up
Keck spectroscopy to study the evolution of the stellar and dark
matter content of early-type galaxies as a function of cosmic time.
\end{abstract}



\section{Introduction}\label{sect:intro}

The last two decades have seen the emergence of a 
hierarchical model for the formation of structure in the
universe. The main ingredients, dominating the overall dynamics of
the universe, are non-relativistic particles that do not interact with
light or baryons except through gravity (i.e., cold dark matter), and
the mysterious dark energy \citep[][]{Per++99,Rie++98}, i.e., a term in
the stress-energy tensor of the universe characterized by negative
pressure (or equivalently a cosmological constant).

The familiar standard model particles move and interact with each
other within this skeleton of unknown particles and fields
\citep{W+R78}. Although standard model particles represent only a small
minority of the total energy budget of the universe \citep{Kom++11},
they play a crucial role in the formation of galaxies, and their
constituents stars and planets. In those overdense regions of the
universe, interactions between standard model particles alter their
dynamics and spatial distribution. Indirectly, through gravitational
interactions, this so-called baryonic physics in turn modifies the
properties of the underlying dark matter distribution
\citep{Blu++86,Gne++04,Duf++10}. Understanding the interplay between 
baryons and dark matter at sub-galactic scales is crucial not only for
any effort to understand how galaxies form and evolve, but also may
shed light on the properties of the dark matter itself
\citep[e.g., self-interaction cross-section;][]{S+S00,L+W11}.

From an observational point of view, it is very challenging to measure
accurately the relative distribution of baryonic and dark matter on
sub-galactic scales. By-and-large, traditional methods require the
presence of a luminous tracer (e.g., hot plasma, cold gas, or stars),
whose kinematics are then interpreted to reconstruct the underlying
gravitational potential
\citep[e.g.,][]{SBS92,Ger++01,Hum++06,Fvd94}.
Gravitational lensing \citep[especially
strong lensing at sub-galactic scales;][]{Tre10} provides an additional
and powerful tool to shed light on dark matter. By exploiting the
deflection of light rays from background sources it need not rely on
the presence of luminous tracers in the deflector. Furthermore,
gravitational lensing is only sensitive to the total gravitational
potential and therefore can provide accurate measurements of mass and
mass distribution independent of its dynamical state or nature.

The combination of strong gravitational lensing with other diagnostic
tools, such as stellar kinematics
\citep[e.g.,][]{Mir95,N+K96,T+K02a,T+K04,K+T03,San++04,Gav05,Koo++06,Koo++09,Gri++08b,Czo++08,Czo++12,Bar++09a,Bar++09b,Bar++11a,Son++12},
weak lensing \citep{Gav++07,Lag++10,Aug++10}, stellar populations
synthesis methods
\citep[e.g.,][]{Gri++09,Aug++09,Aug++10,Tho++11,Tor++10,Tre++10,Spi++11,Spi++12}, etc.,
is particularly effective. By breaking a number of degeneracies
inherent to each method alone one can give precise answers to a number
of questions. What is the relative abundance of dark and luminous
matter in the inner parts of galaxies?  Are dark matter density
profiles universal as those predicted by simulations?  Are halos as
triaxial as predicted by simulations?  How much of the dark matter
observed in galaxies is baryonic (i.e., low-mass stars and high-mass
stars remnants) and how much is non-baryonic?

Until a few years ago, answers to these questions based on strong
gravitational lensing where mostly limited by the small samples of
known lenses.
In fact, strong lensing is a relatively rare phenomenon,
and in general only $\sim1/1000$ background galaxy will have a massive
foreground deflector sufficiently well aligned along the line of sight
to produce multiple images \citep[e.g.,][]{TOG84,SEF92,Chae03}.
However, this situation is changing rapidly owing to
the dedicated efforts of a number of groups in exploiting massive
imaging and spectroscopic surveys such as the Sloan Lens ACS Survey (SLACS),
the Boss Emission Lines Lens Survey (BELLS), the HST Archive Galaxy-scale Gravitational Lens Survey,
or the Strong Lensing Legacy Survey (SL2S) to find samples of strong lenses
\citep{Bro++03,Bol++06,Mar++09,Tre++11,Bro++12}.  
In the past few years, dedicated searches using a variety of techniques
and wavelengths have delivered well over 200 galaxy-scale strong
lenses.

\subsection{The Strong Lensing Legacy Survey}\label{sec:sl2sdef}
The SL2S \citep[][]{Cab++07} is a dedicated
effort to find strong lens systems in the Canada-France-Hawaii
Telescope Legacy Survey (CFHTLS)\footnote{See
\url{http://www.cfht.hawaii.edu/Science/CFHLS/} and links therein for
a comprehensive description} with the goal of answering fundamental
questions about the distribution of mass and light. It consists of two
main efforts: the group and cluster-scale survey
\citep{Lim++09,Mor++12,Ver++11} and this present series that is
concerned with the follow-up and analysis of the galaxy-scale sample
(R. Gavazzi et al., in preparation). 

The candidates are identified using the {\tt RingFinder} algorithm
which detects compact rings around centers of isolated galaxies
($\lesssim 10^{13} M_{\sun}$), and works by looking for blue
features in excess of an early-type galaxies (ETGs) smooth light
distribution that are consistent with the presence of lensed arcs.
After selecting a sample of bright ($i_{\rm AB}
\le 22.5$) red galaxies in the redshift range $0.1\le z \le
0.8$, a scaled, Point Spread Function (PSF) matched version of the $i$-band image was
subtracted from the $g$-band image. The rescaling in this operation is
performed such that the ETG light is efficiently removed, leaving only
objects with a spectral energy distribution different from that
of the target galaxy. These typically blue residuals are then
characterized with an object detector, and analyzed for their
position, ellipticity, and orientation, and those with properties
consistent with lensed arcs are kept as lens candidates. In
practice, we require the blue excess to be elongated (axis ratio
$b/a<1/2$) and tangentially aligned ($\pm 25^\circ$) with respect to
the center of the foreground potential deflector. We also consider
favorably multiple residual objects with similar colors. The objects
are searched for within an annulus of inner and outer radius
$0\farcs5$ and $2\farcs7$, respectively. The lower bound is chosen to
discard fake residual coming from the unresolved inner structure of
the deflector, inaccurate PSFs, etc. The outer bound is chosen to
limit the detection of the many singly imaged objects that only
experience a modest amount of shear (see, R. Gavazzi et al., in preparation,
for further details).  A sample of several hundred good candidates
was visually inspected and ranked for follow-up Hubble Space Telescope (HST) imaging
and Very Large Telescope (VLT) or Keck spectroscopy. These observations are
required to confirm the actual lensing nature of the systems and to
allow accurate lens modeling.  In practice, our selection process
finds about 2-3 lens candidates per square degree. Extensive follow-up
shows that the sample is 50\%-60\% pure (see, R. Gavazzi et al., in preparation).
The SL2S galaxy-scale sample provides an ideal higher-$z$
complement to the SLACS: the deflectors have similar distribution in
size and velocity dispersion, but they have a median redshift of $\sim
0.5$ (cf. $z_d\sim0.2$ for SLACS), extending the baseline for
evolutionary studies back to {two thirds of} the current age of the
universe.

In this series we construct lens models for a pilot sample of
confirmed SL2S galaxy-scale lenses for which Keck spectroscopy is
available, and interpret them by themselves and in combination with
stellar kinematics and stellar population synthesis models
\citep[Paper II;][]{Ruf++11}. This first paper presents lens models for
16 confirmed systems as well as one candidate which proved
impossible to model with simple gravitational potentials and was
therefore discarded as a strong lens.\footnote{A successful
measurement of the redshift of the blue arc-like features might
convince us to consider a more complex lens model.} We use the
results of the lens modeling to discuss relative orientation of mass
and light and their flattening. Following the approach of many lens
surveys before us \citep[e.g.,][]{Bol++06,Bol++08b,Tre++06,Koo++06}, we
carry out this pilot analysis using standard but relatively simple
models, in order to provide an initial benchmark and to illustrate the
quality of the data and the potential of the survey. Specifically, the
models presented here are based on singular isothermal elliptical
potentials and simply parameterized sources. Using the Einstein radii
presented here, the companion Paper II discusses the relative
abundance of mass and light and the evolution of the mass density
profile by combining the SL2S, SLACS, and Lens Structure and Dynamics (LSD) samples.
We are currently working to enlarge the sample substantially and gather more
follow-up data. Future papers of this series will present the enlarged
sample as well as more sophisticated lens models designed to exploit
the richer dataset and sample size.

This paper is organized as follows. In~Section~\ref{sect:data} we
introduce the CFHT and HST data used in the
analysis. In~Section~\ref{sect:models} we describe our modeling
techniques. Section~\ref{sect:results} describes the lens models and
presents our results on the relative alignment of mass and light. A
brief discussion of the properties of the lensed sources is also
presented. Section~\ref{sect:conclude} concludes with a brief summary.

Throughout this paper magnitudes are given in the AB system.  We
assume a concordance cosmology with matter and dark energy density
$\Omega_m=0.3$, $\Omega_{\Lambda}=0.7$, and Hubble constant $H_0$=70
km s$^{-1}$Mpc$^{-1}$.

\section{Observations}\label{sect:data}

\subsection{CFHT Data}\label{ssect:cfht}

The CFHTLS consists of two main components of sufficient depth and
image quality to be interesting for lens searches\footnote{\url{http://www.cfht.hawaii.edu/Science/CFHLS}}.
Both are imaged in the $u^*$, $g$, $r$, $i$ and $z$ bands with the 1~deg$^2$
field-of-view Megacam Camera. The multi epoch Deep Survey covers four
pointing of 1~deg$^2$ each. Two different image stacks were produced:
D-85 contains the 85\% best seeing images whereas the D-25 only
includes the 25\% best seeing images. For finding lenses we only
considered the better resolution stacks. In the T06 data release \citep{Gor++09}
\footnote{See \url{http://terapix.iap.fr/cplt/T0006/T0006-doc.pdf}} used
for this study they reach a typical depth of $u^*\simeq 26.18$,
$g=\simeq 25.96$, $r\simeq 25.43$, $i\simeq 25.08$,
and $z\simeq 24.57$ (80\% completeness for point sources) with
typical FWHM PSFs of $0\farcs75$, $0\farcs69$,
$0\farcs64$, $0\farcs62$ and $0\farcs61$, respectively. The Wide
survey is a single epoch imaging survey, covering approximately 171 deg$^2$ in
4 patches of the sky. It reaches a typical depth of $\rm u^*\simeq
25.35$, $\rm g\simeq 25.47$, $\rm r\simeq24.83$, $\rm i\simeq 24.48$,
and $\rm z\simeq 23.60$ (AB mag of 80\% completeness limit for point
sources) with typical FWHM PSFs of $0\farcs85$,
$0\farcs79$, $0\farcs71$, $0\farcs64$ and $0\farcs68$, respectively.
Owing to the greater survey solid angle, the Wide component is our main source
of lens candidates.

\begin{figure*}
  \centering
  \includegraphics[width=0.90\linewidth]{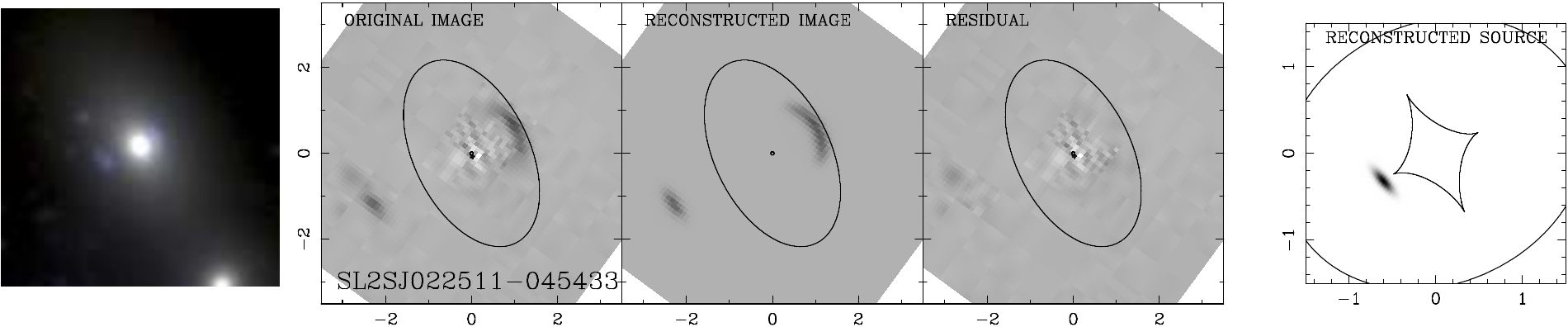}
  \includegraphics[width=0.90\linewidth]{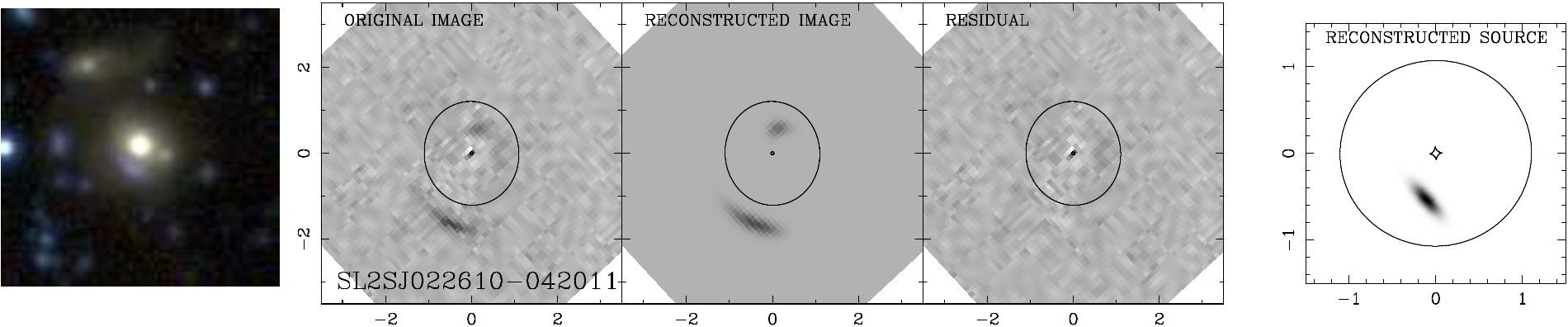}
  \includegraphics[width=0.90\linewidth]{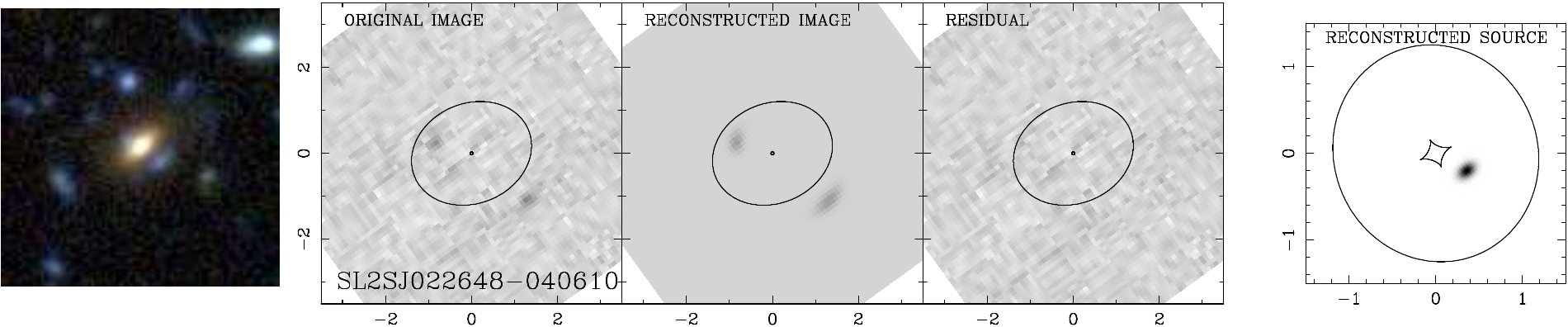}
  \includegraphics[width=0.90\linewidth]{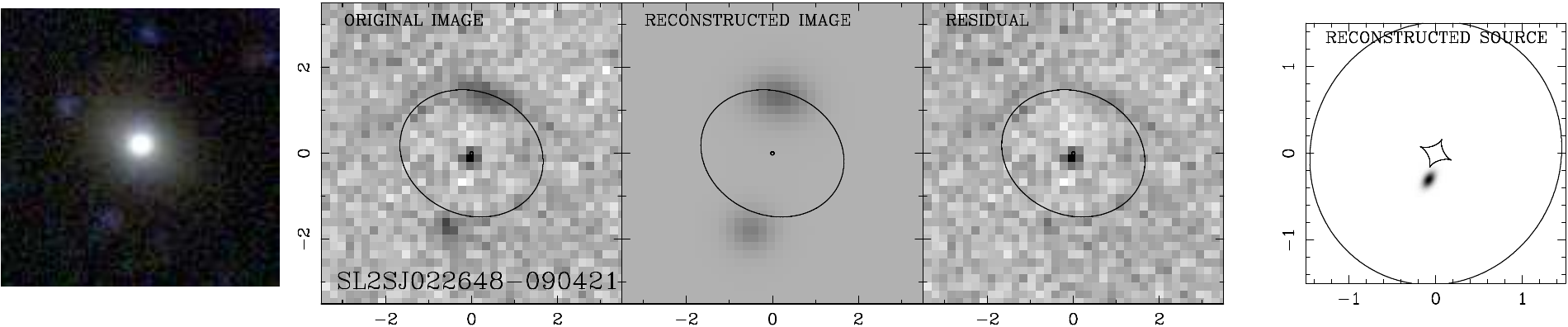}
  \caption{\label{fig:models2} Lens models (continued).}
\end{figure*}

\begin{figure*}
  \centering
  \includegraphics[width=0.90\linewidth]{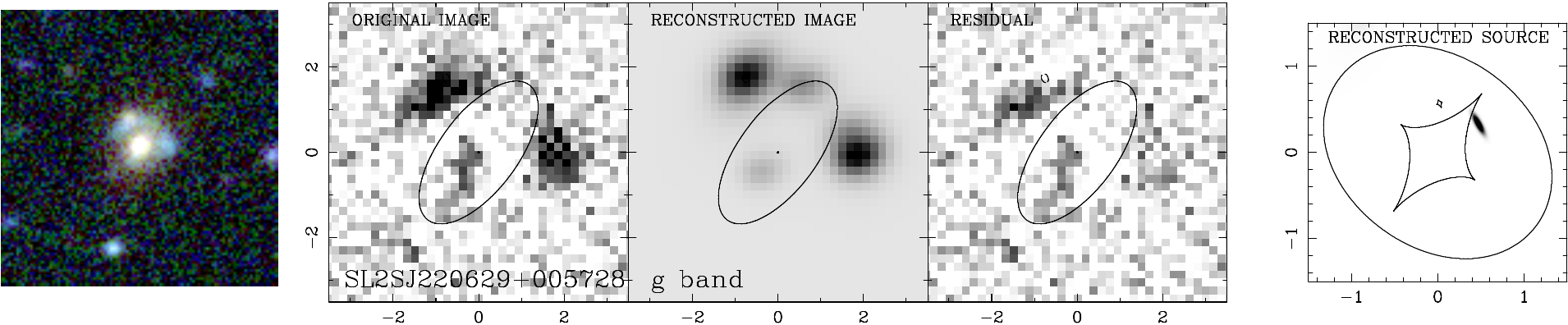}
  \includegraphics[width=0.90\linewidth]{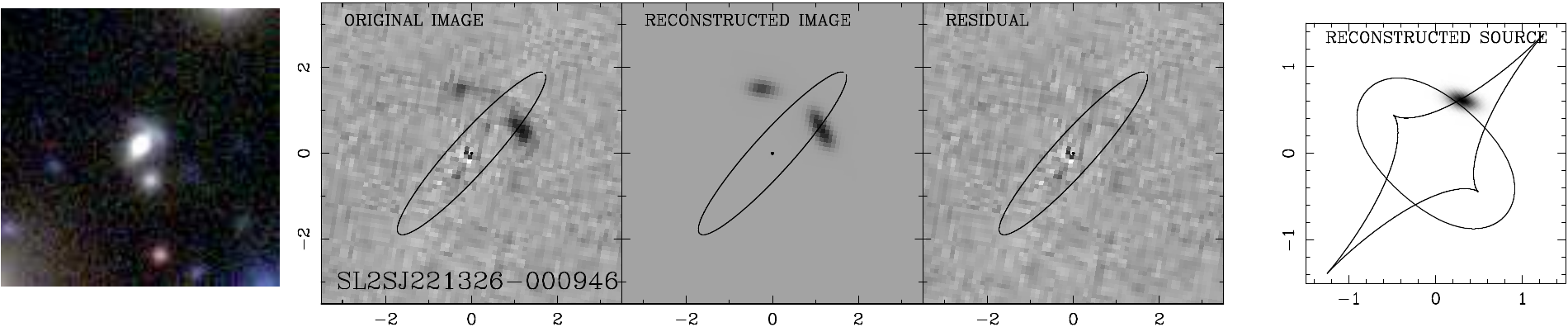}
  \includegraphics[width=0.90\linewidth]{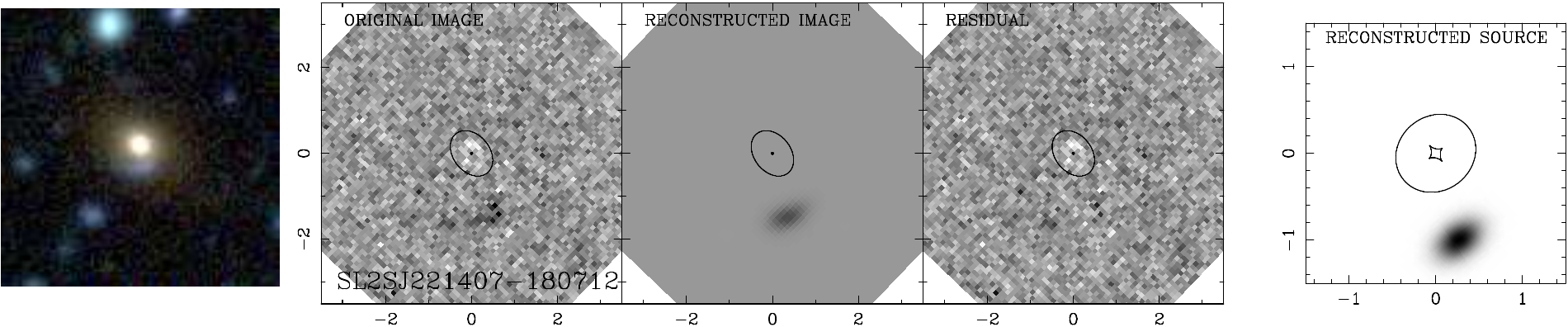}
  \includegraphics[width=0.90\linewidth]{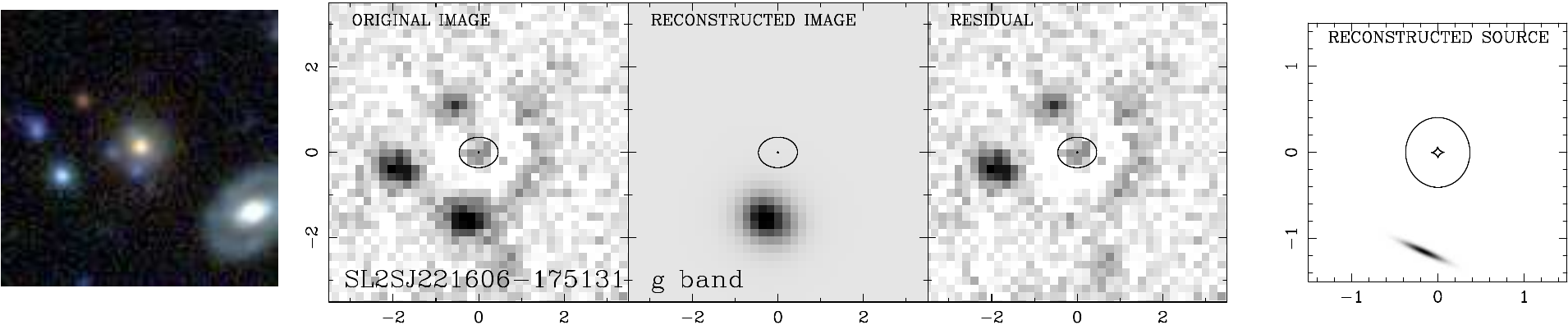}
  \includegraphics[width=0.90\linewidth]{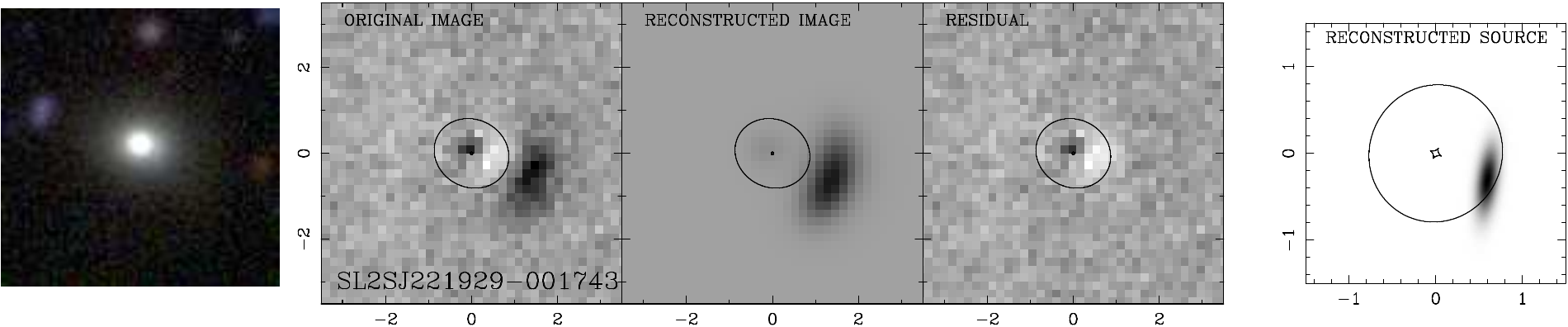}
  \caption{\label{fig:models4} Lens models (continued). Note that the system
in the third row, SL2SJ221606-175131, is not a satisfying model of a lens and
we disqualify it as being an actual gravitational lens.}
\end{figure*}

Around each candidate lens and in each CFHTLS band, we produce cutout
images 101 pixels (\ie~$18\farcs8$) wide. Nearby stars within 5
arcmin from a given lens are used to produce a PSF model. The 17
systems analyzed in this study are shown in
Figures~\ref{fig:models1}-\ref{fig:models4} and listed in
Table~\ref{table:main}.

\subsection{{\it HST} Follow-up Imaging}\label{ssect:hst}

In order to confirm the lensing hypothesis and allow for detailed lens
modeling, 65 galaxy-scale lens candidates have been observed with the
HST as snapshot programs during cycles 15,
16, 17 (GOs 10876, 11289, PI: J.-P. Kneib; GO 11588, PI: R. Gavazzi). The
observations started with the Advanced Camera for Surveys (ACS), then
switched to the Wide Field and Planetary Camera 2 (WFPC2) after the
failure of ACS, and finally turned to the Wide Field Camera 3 (WFC3)
after Servicing Mission 4.  Approximately, 50\% of the lens candidates
were confirmed as lenses in this way.  A more comprehensive
description of the efficiency of the SL2S lens finding strategy will
be given by R. Gavazzi et al. (in preparation).

Ten of the galaxy-scale systems were observed with {\em HST} early enough to
be included in our first Keck spectroscopic follow-up campaign and are
the subject of papers I and II. The remaining systems will be
presented at the end of our Keck and HST observing campaigns in the
next papers of this series. Of these ten systems, three were observed
in two bands, with ACS or WFC3 whereas the others were observed only
in a single WFPC2 band as detailed in Table~\ref{table:hst}.

All the WFPC2 data were reduced using standard {\tt
MultiDrizzle}\footnote{{\tt MultiDrizzle} is a product of the Space
Telescope Science Institute, which is operated by AURA for NASA
\citep{Koe++02}.} recipes. The cosmic ray removal worked well because
the 1200 s exposure time was split into three exposures. The
``drizzling'' was performed by preserving the CCD frame orientation
and pixel scale to avoid producing correlated noise. The ACS and WFC3
observations consisted of single or double exposures only and
therefore we relied on {\tt LA-Cosmic} \citep{vDo01} on individual
exposures for cosmic ray removal, before combining them with {\tt
swarp} \citep{Ber++02}. As for WFPC2, images are kept in the natural
CCD frame.  Exposure times and final pixel scales are given in
Table~\ref{table:hst}.

\begin{deluxetable}{lrrc}
\tablecaption{ \label{table:hst} Summary of HST observations.}
\tablehead{
name & instrument/filter & exp. time & pixel scale\\
     &   & (sec) & (arcsec) }
\startdata
SL2SJ021411$-$040502  &  ACS/F606W   &  400 & 0.05 \\
                      &  ACS/F814W   &  800 & 0.05 \\
SL2SJ021737$-$051329  &  ACS/F606W   &  400 & 0.05 \\
                      &  ACS/F814W   &  800 & 0.05 \\
SL2SJ022511$-$045433  &  WFPC2/F606W & 1200 & 0.1 \\
SL2SJ022610$-$042011  &  WFPC2/F606W & 1200 & 0.1 \\
SL2SJ022648$-$040610  &  WFPC2/F606W & 1200 & 0.1 \\
SL2SJ023251$-$040823  &  WFPC2/F606W & 1200 & 0.1 \\
SL2SJ140123$+$555705  &  WFPC2/F606W & 1200 & 0.1 \\
SL2SJ141137$+$565119  &  WFC3/F475X  &  720 & 0.04 \\
                      &  WFC3/F600LP &  720 & 0.04 \\
SL2SJ221326$-$000946  &  WFPC2/F606W & 1200 & 0.1 \\
SL2SJ221407$-$180712  &  WFPC2/F606W & 1200 & 0.1 \\
\enddata
\end{deluxetable}

\section{Modeling methodology}\label{sect:models}

In order to model the light distribution of a galaxy-scale
gravitational lens, one has to disentangle the contribution of the
foreground deflector and that of the background lensed arc-like
features. The former is generally a red ETG and the former a dimmer,
blue, and presumably star-forming, more distant source. The light
distribution of an ETG generally has a regular shape sufficiently well
described by a S\'ersic \citep{Ser68} or even a de Vaucouleurs profile
\citep{deV48} with very small color gradients. Blue background
sources are also well represented by Exponential
profiles\footnote{More complex models involving pixelated source will
be explored in the next papers of this series with a large sample and
complete HST follow-up (A. Sonnenfeld et al., in preparation).}
\citep{New++11}.  The separation of these two components is relatively
straightforward for ETGs deflectors and they can generally be fitted
independently for most applications. However, very detailed
investigations of the source properties of high signal-to-noise data
might require a simultaneous fit of both components
\citep[e.g.,][]{Mar++07}, especially with ground-based data (F. Brault et
al., in preparation) and bright arcs. We describe our procedure to fit the
foreground deflector in Section~\ref{ssect:moddef}.
In Section~\ref{ssect:modearc} we describe 
how we use the residual lensed images to model the gravitational
potential of the deflector and the intrinsic surface brightness of the
source.

\subsection{Foreground Deflector}\label{ssect:moddef}

\begin{figure*}
  \centering
  \includegraphics[width=0.90\linewidth]{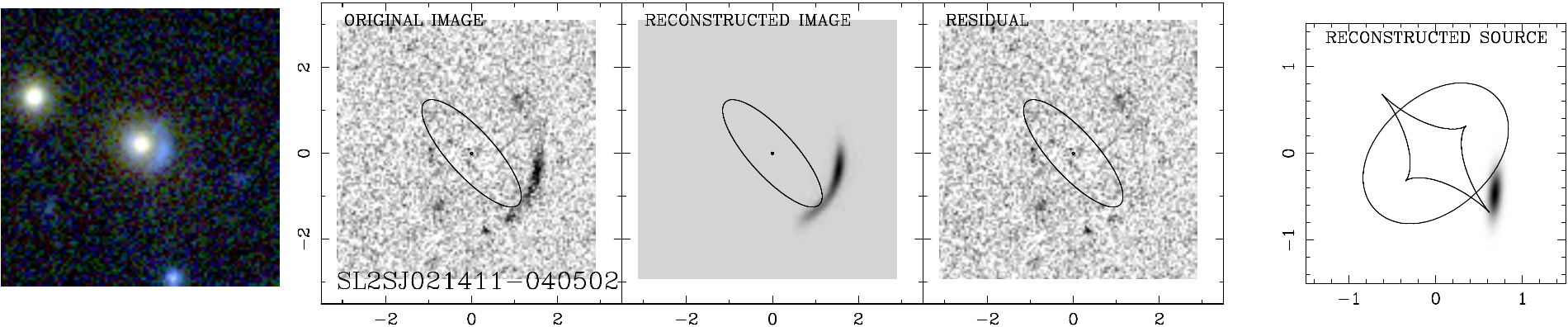}
  \includegraphics[width=0.90\linewidth]{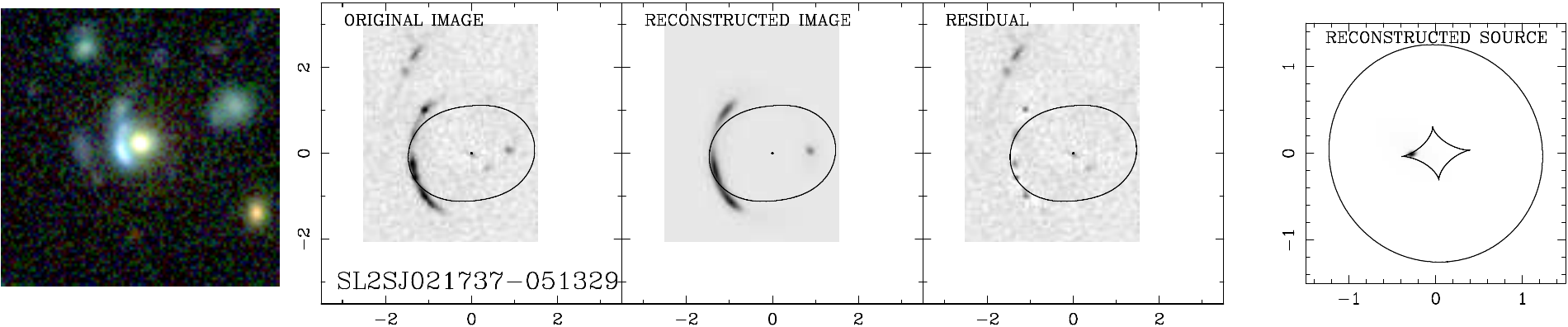}
  \includegraphics[width=0.90\linewidth]{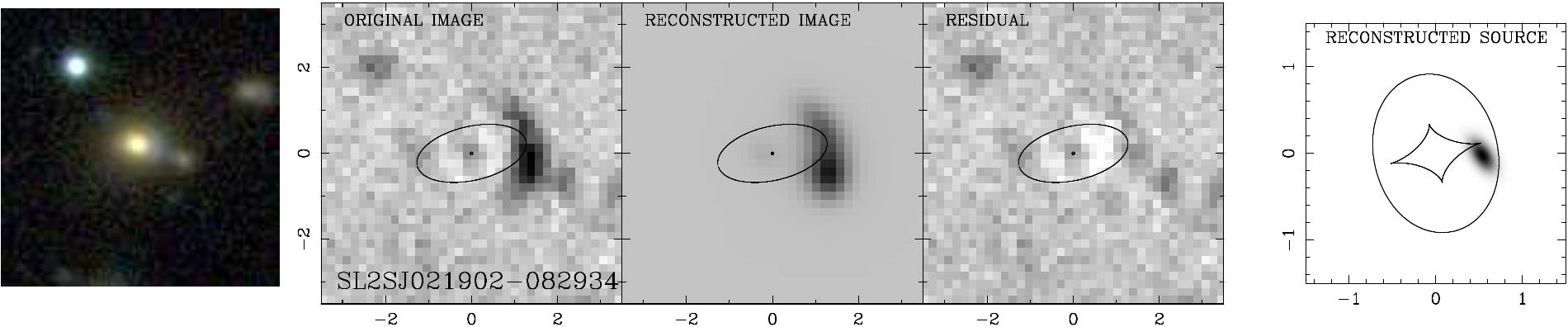}
  \includegraphics[width=0.90\linewidth]{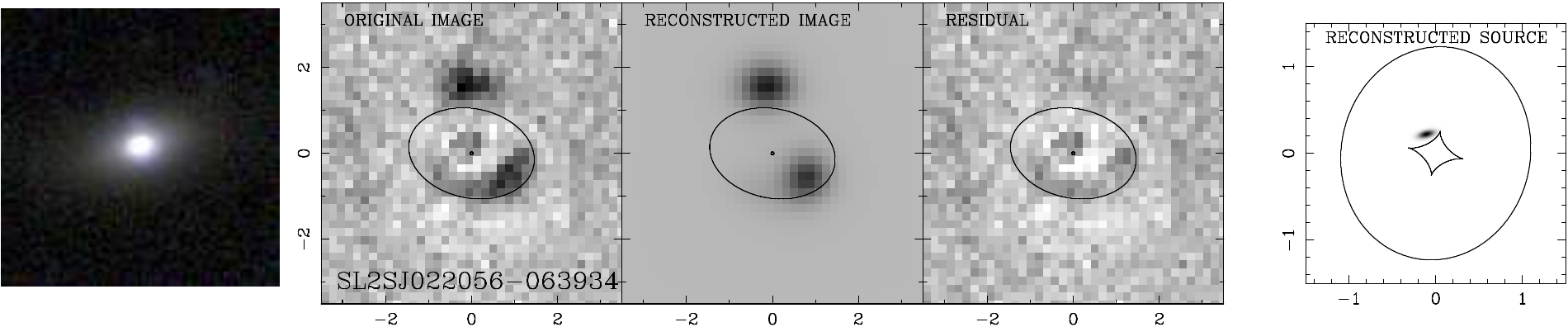}
  \caption{\label{fig:models1} Results of the lens modeling analysis. For each row, a system is shown with, from left to right, a $18\farcs8\times18\farcs8$ CFHTLS gri composite image of the system, the lensed features with the deflector subtracted off with {\tt galfit}  (data), the best-fit model prediction (model), the residual (data$-$model) and the associated source plane light distribution. The critical lines are overlaid in the first three panels whereas the caustics lines are shown in the last one.}
\end{figure*}

We used the versatile {\tt galfit} software \citep{Pen++02} to perform
the subtraction of the foreground deflector as it allows to account
for boxyness/diskyness of isophotes. It generally yields good image
subtraction with a S\'ersic profile (see Figure~\ref{fig:models1}) but
the recovered S\'ersic indices $n$ and effective radii $R_{\rm eff}$
are quite degenerate. 

We thus used a generic S\'ersic profile with more degrees of
freedom to get as good a deflector subtraction as possible and we also
applied a strong $n=4$ prior to get a more precise value of $R_{\rm
eff}$. In the same vein, the fit of the $n=4$ foreground lens also
yields a robust measurement of ellipticity and orientation of the
light distribution. These values are reported in
Table~\ref{table:main}.

Fits are performed in all the available bands of a given lens using
$18\farcs8$ on a side cutout images and suitable PSF either inferred
from the neighboring stars or from {\tt TinyTim}
\citep{Kri++11}\footnote{\url{http://www.stsci.edu/hst/observatory/focus/TinyTim}}
in the case of {\em HST} data. The formal errors on each parameter are 
generally very small because they only account for statistical errors
and not modeling errors associated with the mismatches between the
form of the assumed and observed light distribution. A more realistic
estimate of the total uncertainties on the recovered ellipticity,
orientation and effective radius can be made by estimating the
filter-to-filter dispersion on these parameters when {\em HST} data are
missing. When {\em HST} imaging is available, shape parameters characterizing the
deflector are more robustly measured and we adopt the formal errors
from {\tt galfit}.

\subsection{Lensed Features and Mass Modeling}\label{ssect:modearc}

For lens modeling we used a dedicated code {\tt sl\_fit} developed for
and tested on galaxy-scale strong lenses
\citep[e.g.,][]{Gav++07,Gav++08,Gav++11,Tu++09}. The code fits model
parameters of simple analytic lensing potentials. It uses the full
surface brightness distribution observed in the image plane and
attempts to explain it with one or more simple analytic light
components described by an exponential radial profile with elliptical
shape \citep[see, e.g.,][for similar techniques]{Mar++07,Bol++08a,New++11}.

The lensing potential is assumed to be made of a singular isothermal
ellipsoid (SIE), centered on the main deflector. This is the simplest
mass profile that has been shown to yield a description of the mass
distribution of massive ETGs sufficient to derive
Einstein radius, position angle, and ellipticity
\citep[e.g.,][]{RKK03,R+K05,Koo++06,Gav++07,Koo++09b,Bar++11a}.  The
convergence profile of the central mass component is given by
\begin{equation}\label{eq:main}
\kappa(x,y) = \frac{b}{2 \xi},
\end{equation}
where the scaling parameter $b$ is the Einstein radius $R_{\rm Ein}$
\citep[e.g.,][]{SEF92,KSB94}. $b$ is related to the velocity
dispersion of the deflector through $b/1\arcsec = (\sigma_v/186.21
\,\kms)^2 D_{\rm ls}/D_{\rm s}$, where $D_{\rm ls}$ and $D_{\rm s}$ are
angular diameter distance between the lens and the source and between
the observer and the source, respectively.  $\xi^2= q x^2+y^2/q$ is
the radial coordinate that accounts for the ellipsoidal symmetry of
the isodensity contours and $q$ is the minor-to-major axis ratio. The
orientation of the major axis P.A.$_{\rm tot}$ is allowed to vary,
although this is not explicit in the definition of
Equation~(\ref{eq:main}). We do not assume a priori that the
orientation of the total mass distribution is correlated to that of
the observed stellar component. In this way, the total potential can
capture (and mix because of substantial degeneracies) internal and
external quadrupolar terms in the potential \citep{KKS97}. We also add
an external shear parameter when more information is available (see
the case of SL2SJ021737$-$051329 described below), sufficient to break
the degeneracy between external shear and orientation of the galaxy
potential.

As in the previous step, when we fitted the foreground light
distribution, we weigh pixels with the image inverse total variance
(including sky, foreground, and lensed features).  The $\chi^2$ term
relating the observed light distribution $I(\mathbf{x}_i)$ at pixel $i$
and the intrinsic source light distribution $S(\mathbf{y})$ is thus:
\begin{equation}\label{eq:chi2def}
\chi^2 = \sum_i \frac{\left[I(\mathbf{x}_i) - S( \mathbf{x}_i - \boldsymbol{\alpha}(\mathbf{x_i} \vert \mathbf{p}_{\rm p} ) \vert  \mathbf{p}_s )\right]^2 }{\sigma_i^2} \; ,
\end{equation}
where $\mathbf{p}_{\rm p}=[b, q, {\rm P.A.}_{\rm SIE}]$ contains the
parameters that determine the gravitational potential and
$\mathbf{p}_s=[x_{\rm s}, y_{\rm s}, q_{\rm s},{\rm P.A.}_{\rm s}, m_{\rm
s}, r_{\rm s}]$ is the section of the parameter space that describes
the source light distribution, namely, its center $\mathbf{x}_s$, its
ellipticity $q_{\rm s}$, its orientation P.A.$_{\rm s}$, its magnitude
$m_{\rm s}$, and half-light radius $r_{\rm s}$.
 
The optimization of these nine parameters is performed using Monte Carlo
Markov Chain techniques.  Reported model parameters and confidence
intervals are taken from the 0.50, 0.16, and 0.84 quantiles. However,
because of the simplicity of both the model potential and model source
light distribution, we generally end up with very small formal errors
on the recovered model parameters that should be substantially
increased. \citet{Bol++08a} estimated that relative errors on $R_{\rm
Ein}$ should be about 5\%. We thus add this dominant contribution in
quadrature to the statistical errors in Table~\ref{table:main} and do
the same for the axis ratio $q_{\rm SIE}$ by adding a 
$\pm0.07$ absolute error on consistent with \citet{New++11} and even
more conservative than the $\pm0.05$ error quoted by \citet{Bol++08a}.

\section{Results}\label{sect:results}

In this section we describe the lens model of each system (Section~4.1). In
addition, we study the shape and relative orientation of the light and
total mass distributions as they do are independent of the source and
deflector redshift (Section~4.2). Results depending on spectroscopic
information (source and deflector redshifts and velocity dispersion)
along with a novel method proposed to mitigate the nuisance
due to the ignorance of the source redshift are presented in Paper II.

\begin{figure*}
  \centering
  \includegraphics[width=0.90\linewidth]{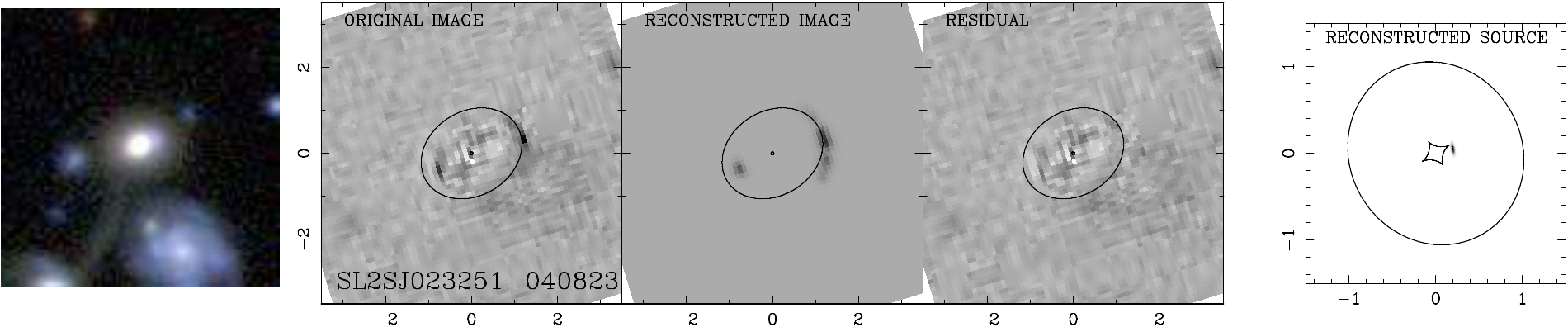}
  \includegraphics[width=0.90\linewidth]{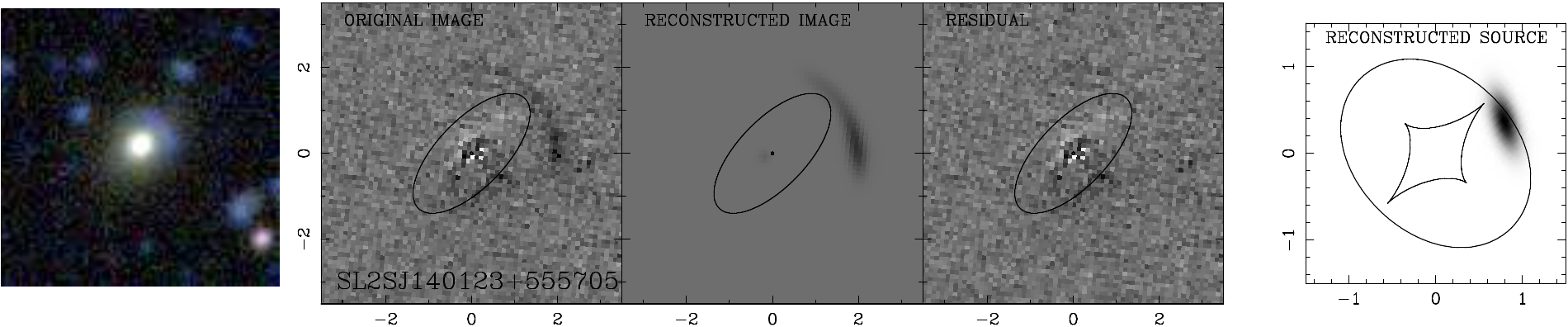}
  \includegraphics[width=0.90\linewidth]{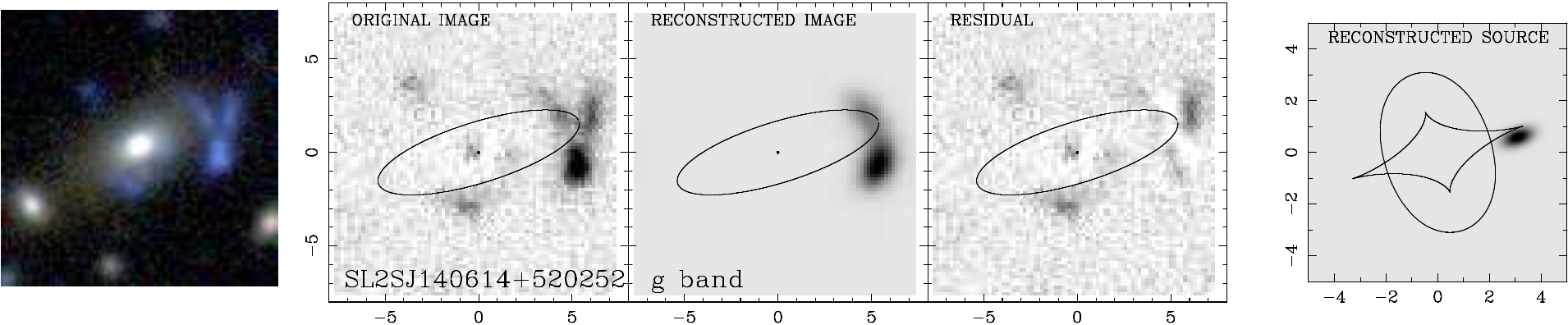}
  \includegraphics[width=0.90\linewidth]{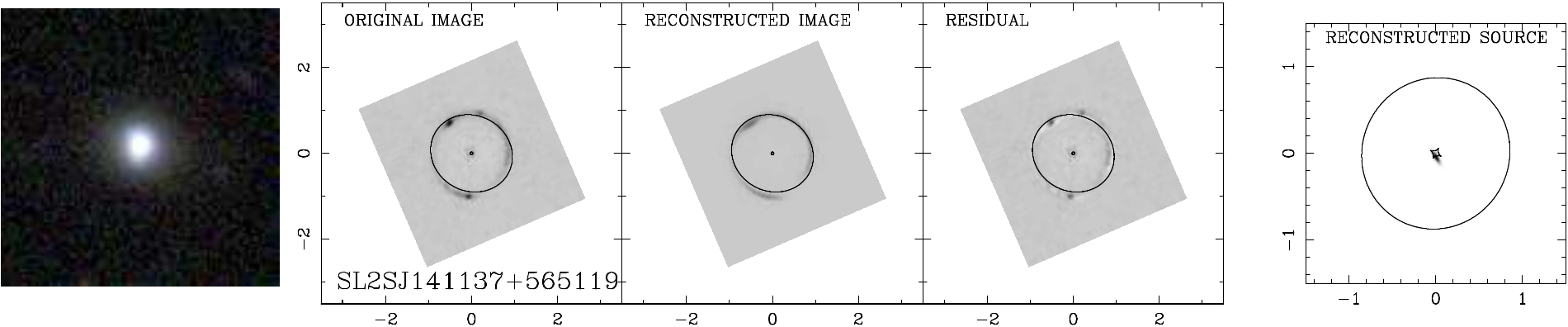}
  \caption{\label{fig:models3} Lens models (continued). Note the different plotting range for the third row corresponding to SL2SJ140614+520252.}
\end{figure*}

\subsection{Notes on Individual Lens Models}

For each lens in the sample we describe the resulting best-fit model.
\begin{itemize}
\item SL2SJ021411$-$040502 can be reproduced by a single SIE potential + Exponential source model, located just outside the caustic. We note a small axis ratio $q_{\rm SIE}\sim 0.33$ this seems at odds with the more
circular light distribution $q_* \sim 0.89$. This can be explained by
the presence of a neighbor galaxy at the same redshift as $z_{\rm d}$
about 8\arcsec and ${\rm P.A.}\sim65^\circ$ which enhances the elongation of
the potential. An alternative solution with a slightly lower
likelihood is found where the source is located just inside the
caustic and the lensing configuration consists of three merging images
(one of which is barely visible in the {\em HST} image). This solution has a
larger Einstein radius (by $\sim$30\%), a slightly rounder potential,
and a different position angle (elongation almost east--west). For the
present study we adopt the solution with the highest likelihood
although the secondary solution cannot be excluded completely without
deeper {\em HST} data.
\item SL2SJ021737$-$051329 has been studied by in detail by
\citet{Tu++09}. Our results are in agreement with theirs. A simple SIE
+ Exponential source yields a good fit, even though a better fit is
achieved by introducing both internal ellipticity through the SIE
potential and external shear (consistent with the presence of a nearby
group of galaxies).  Similar conclusions were recently found by
\citet{Coo++11} using additional near-IR CANDELS data. The source
redshift is $z_{\rm s}=1.847$ (Paper II). This implies that the source
half-light radius is $R_{\rm eff,s}=0\farcs081\mypm{0.028}{0.016}
\simeq 0.69 \mypm{0.24}{0.13}\,\kpc$ for an F606W magnitude $m_{\rm s}
= 24.45 \pm 0.10$. The system is lensing a second source at
$z_{s2}\sim 2.3$ that we do not consider here for consistency with the
other systems.  However, regardless of whether the second source is
modeled or not, the good signal-to-noise ratio and the favorable
image configuration allows us to break the degeneracy between external
shear and internal ellipticity and we therefore get constraints on
both. We find an external shear $\gamma_{\rm ext}=0.064 \pm 0.003$
with an orientation P.A.$_{\rm ext}=89\fdg7\pm 0\fdg4$. We see that the
light is well aligned with the external shear and about $\sim 33\deg$
misaligned with the SIE component. The SIE mass component is found
to have a rather circular distribution ($q_{\rm SIE}=0.91\pm0.07$, Table \ref{table:main}),
we thus do not consider this apparently strange behavior too seriously as some of the mass
ellipticity (internal quadrupole) could be exchanged with external shear with a slightly
different mass profile \citep{Tu++09}.
\item SL2SJ021902$-$082934 is another "cusp" configuration with a
marginal candidate counter image on the opposite side. The lack of {\em HST}
imaging implies that we had to consider the CFHT $g$-band image. We
can see an additional source westward of the arc that has a color
similar of the deflector and should not be considered as a lensed
feature. Since we can achieve a good fit of the arc without
introducing this perturbation we neglect it.
\item SL2SJ022056$-$063934 is a minor axis cusp configuration. The
single source component does not perfectly capture the faint tail of
the inner arc although most of the flux is well recovered, even with
ground-based resolution.
\item SL2SJ022511$-$045433 is a low-redshift bright deflector with
another bright minor axis cups configuration. The source redshift is
$z_{\rm s}=1.199$ (Paper II). We report here the result with a single
source component of magnitude $m_s=24.14\pm0.04$ in the F606W band and
half-light radius $R_{\rm eff,s}= 0\farcs125 \pm 0\farcs 003 \simeq
1.037 \pm 0.025 \,\kpc$. Note that we also attempted to account for the
fainter extension of the furthest arc (on the east of the deflector),
with a secondary component. This would even lower the residuals on the
two multiple images without changing the results on the recovered
potential parameters. Accounting for this component the source would
be $\sim0.36$ mag brighter.
\item SL2SJ022610$-$042011 is a typical large impact parameter double
configuration implying a substantial differential magnification of the
two multiple images. In addition, these two are nearly aligned with
the center of the deflector. This is consistent with the potential
being close to circularly symmetric ($q_{\rm SIE}\sim 0.92$). This
system has a known source redshift $z_{\rm s}=1.232$. We can thus
estimate the source half-light radius $R_{\rm eff,s}= 0\farcs160 \pm
0\farcs 011 \simeq  1.332 \pm 0.092 \,\kpc$ for an F606W  magnitude
$m_{\rm s} = 25.10 \pm 0.08$.
\item SL2SJ022648$-$040610 is another double configuration with a more
balanced magnification ratio between the two images involving a
slightly more elongated potential. We note that the deflector looks
like an edge-on S0 galaxy with an elongated light
distribution. However, the total potential is considerably more
circular.
\item SL2SJ022648$-$090421 is also modeled using Megacam $g$ band data
and shows a minor axis cusp configuration with little deviation
between light and SIE orientations. This is the faintest source we
reconstruct with $m_{\rm s}=27.13\pm0.14$.
\item SL2SJ023251$-$040823 is a double system with the source close to the cusp.
\item SL2SJ140123+555705 shows a dim cusp-like arc. Even though we
cannot identify the counterimage, the substantial bending of the arc
breaks the degeneracy between shear and intrinsic source ellipticity
and allows us to measure the Einstein radius with good accuracy
$R_{\rm Ein}=1\farcs 19 \pm 0\farcs07$, comparable to the other cases.
\item SL2SJ140614+520252 is another cusp configuration on a larger
scale, as $R_{\rm Ein}=3\farcs02 \pm 0\farcs15$. Some blue light
excesses are seen in the $g$-band image shown in
Figure~\ref{fig:models3} and only the main ``naked'' cusp arc is
captured by our model. However, we can see that the other western
component is not multiply imaged as the source would fall outside the
caustic. Thus, it does not provide more information and we verified
that its inclusion does not change our conclusions on the lensing
potential. Even by ignoring this second singly imaged source
component, we find that this system involves the brightest source of
the sample with $g$ band $m_{\rm s}=23.50\pm 0.03$. We could not find
a solution with a lensed source reproducing the southern and
northeastern blue excesses. The lens potential is highly elongated
with $q_{\rm SIE}=0.290\pm 0.015$.
This is probably due to the additional contribution of a nearby galaxy about $8\farcs3$ in the southeast direction that corresponds to the major axis of the already quite elongated light component.
\item SL2SJ141137+565119 is the only system having WFC3 imaging
presented here.  The signal-to-noise ratio is good and allows us to
get tight constraints on the potential and on the source whose
redshift is found to be $z_{\rm s}=1.420$. The source half-light
radius in the F475X band is $R_{\rm eff,s}= 0\farcs0.058 \pm
0\farcs001 \simeq 0.490 \pm 0.008 \,\kpc$ and its magnitude is
$m_{\rm s}=25.86\pm 0.04$. We note that the addition
of a very compact core would improve the fit further as our single
source component does not reproduce the full complexity of the source
light distribution. Our best-fit model nevertheless captures most of
the extended Einstein ring.
\item SL2SJ220629+005728 is modeled using the $g$-band Megacam
image. However, we see in the color cutout image that a small red
satellite lies on top of the lensed blue features. We thus need to
disentangle both contributions by simultaneously fitting to the $g$
and $i$ bands one lensed source and one foreground unlensed source
centered in the red satellite. This satellite also contributes to the
lens potential through a point mass component of unknown mass. We find
that this perturbing mass has to be $M_{\rm pert} \le 2 \times 10^9
\,\msun$ (68\% CL). Getting {\em HST} imaging data would significantly improve the
accuracy of the decomposition and the constraints on $M_{\rm
pert}$. The Einstein radius estimate is robust with respect to the
inclusion (or not) of the perturbing potential.
\item SL2SJ221326$-$000946 is an edge-on disk galaxy in which both the
potential and the light distribution are highly elongated: $q_{\rm
SIE}= 0.191 \pm  0.014 $ and $q_*= 0.277 \pm  0.007$ with a small
$\Delta {\rm P.A.} \sim 10\fdg4 \pm 0\fdg6$ misalignment.
\item SL2SJ221407$-$180712 does not exhibit any counterimage nor a
strong curvature of the main outer image. This implies that the model
cannot be constrained well and we can only place upper limits on the
Einstein radius $R_{\rm Ein} = 0\farcs41 \pm 0\farcs23$. The results
on the potential elongation and orientation are thus very loose and we
do not consider this system in the statistical analysis of
Section~\ref{sec:defstat}. Note that the small upper limit on $R_{\rm Ein}$
is consistent with other mass constraints in Paper II.
\item SL2SJ221606$-$175131 has a relatively symmetric configuration. Its
similarity to a classic quad configuration lead us to include it in
the lens candidate sample. However, we could not find any sensible
lens model able to reproduce in detail the image
configuration. Therefore we cannot conclude that this system is a
lens, or more precisely, that the blue features about 1$\arcsec$ from
the center of the ETG originate from a unique background source. We
speculate that it could be due to low surface brightness star
formation at the ETG's redshift. We thus exclude this system from the
statistical analysis below, although better imaging data would be
useful to revisit this system by constructing more complex models of
the source and potential.
\item SL2SJ221929$-$001743 does not have {\em HST} imaging and the CFHT data
do not allow us to constraint tightly the potential parameters as we
cannot identify unambiguously a counterimage. In addition,  the arc
does not display a strong curvature at the resolution of ground-based
imaging. We thus get weaker constraints on the Einstein radius $R_{\rm
Ein} = 0\farcs74 \pm 0\farcs14$ than for other systems. {\em HST} imaging
would significantly improve the mass model. Given the redshift of the
source $z_{\rm s}= 1.023$, we infer a source half-light radius $R_{\rm
eff,s}=0\farcs384\mypm{0.073}{0.054} \simeq 3.03
\mypm{0.59}{0.43}\,\kpc$ and a $g$ band intrinsic magnitude $m_{\rm
s}=24.08 \pm  0.21$. 
\end{itemize}

\subsection{Alignment of Mass and Light}\label{sec:defstat}

\begin{figure}
  \centering
  \includegraphics[width=0.95\linewidth]{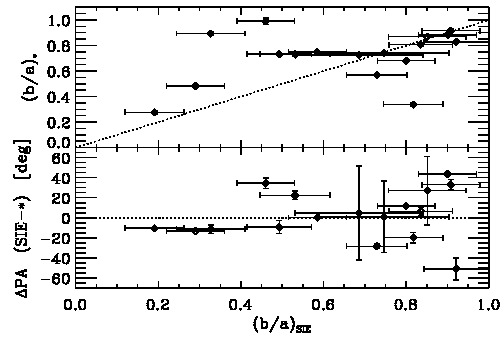}
  \caption{\label{fig:dpa} Top panel: correlation between axis ratio
of the light and of the mass distribution. Bottom panel: angular
offset between the major axis of the mass distribution and that of the
light as a function of axis ratio of the mass
distribution. SL2SJ021737$-$051329 is not included because this system
was modeled including external shear.}
\end{figure}

\begin{figure}
  \centering \includegraphics[width=0.85\linewidth]{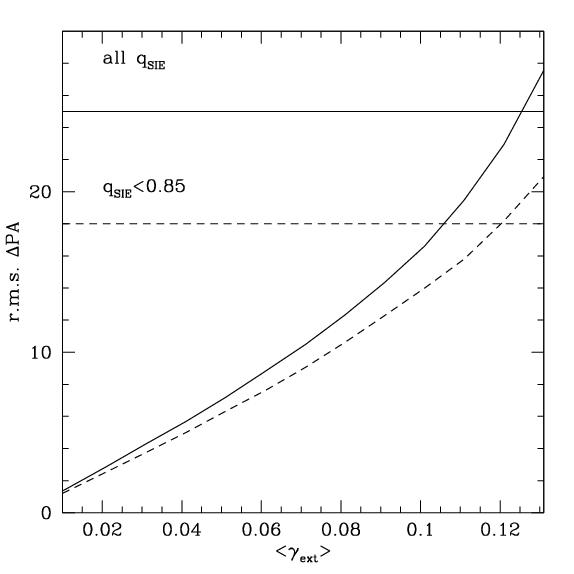}
  \caption{\label{fig:gext} The curves represent the expected rms
  fluctuations of the offset in P.A. as a function of average external
  shear. The horizontal lines represent the values measured for the
  SL2S sample. The solid lines refer to the whole sample, while the
  dashed lines refer to the subsample with mass axis ratio $q_{\rm
  SIE} <0.85$.}
\end{figure}

The top panel of Figure~\ref{fig:dpa} compares the ellipticity of the
stellar component to that of the mass model. For the vast majority of
the objects the ellipticities are tightly correlated. This is expected
since stellar mass makes up a significant fraction of the total mass
within the Einstein radius. However, there are few interesting
outliers. The flattened light distribution of SL2SJ022648$-$040610 is
consistent with a disky galaxy living in significantly rounder
halo. Perhaps more surprising are the cases of SL2SJ021411$-$040502 and
SL2SJ220629+005728, where the stars are almost round, while the
potential is significantly flattened. In both cases the source of the
ellipticity of the mass distribution seems to be external shear,
associated with a satellite. In general, the relation between light
and mass ellipticity seems to have significantly more scatter than
that found for the SLACS sample, $\langle q_{\rm SIE}/q_*\rangle=0.95$ with scatter
0.48, as opposed to 0.99 with scatter 0.11 \citep{Koo++06}. As we
discuss in the next paragraph this is probably due to a much more
significant role played by external shear.
The bottom panel of Figure~\ref{fig:dpa} shows the angular offset
between light and mass P.A. as a function of mass ellipticity. Even
neglecting the points with $q_{\rm SIE}>0.85$, where clearly the
position angle is not very well measured since the potential is so
circular, there is considerable scatter around zero. For the entire
sample the rms scatter is 25 deg, while if we limit ourselves to
$q_{\rm SIE}<0.85$ the scatter is still 18 deg. Again this is
considerable higher than the 10 deg found for the SLACS sample
\citep{Koo++06} in general, and closer to the values found for the
subset of SLACS lenses that reside in overdense environments
\citep{Tre++09}. A simple calculation, based on Equation~(22) in the paper by
\citet{KKS97} and assuming that the direction of external shear is
randomly distributed with respect to that of the mass of the galaxies
shows that this is consistent with a relatively large average external
shear. This is illustrated in Figure~\ref{fig:gext} where we show the
expected rms fluctuation of the position angles as a function of
average external shear. Regardless of whether the rounder objects are
included or not, it seems that external shear $\langle\gamma_{\rm
ext}\rangle= 0.12 \pm 0.05$ is required on average. The mean and
error on this quantity are estimated by propagating the sampling
variance associated with the measurement of the $\pm 25^\circ$ rms
dispersion in the light-mass misalignment angle. The small sample size
prevents us from further estimations like the scatter about this mean
value which, in turn, will be addressed in a forthcoming paper
(Sonnenfeld et al., in preparation).
This level of external shear is fairly common among
galaxy-scale gravitational lenses
\citep[e.g.,][]{KKS97,H+S03}, and it is likely due to the environment
of the lenses
\citep{Aug++07,Tre++09,Aug08,Won++11}, since massive ETGs
typically reside at the centers of groups. The lower level of
external shear in the SLACS sample, $\langle\gamma_{\rm
ext}\rangle\lesssim 0.035$ \citep{Koo++06}, is most likely due to the
smaller size of their Einstein radius relative to the characteristic
scale of the galaxy (half effective radius typically), and therefore
the more relevant role played by stellar mass in defining the
potential within the critical curve. It is also possible that
part of the observed misalignments between the light and the
gravitational potential might arising from intrinsic misalignments
between the galaxy and its halo. This is also expected to be a larger
effect in SL2S than in SLACS, owing to the larger size of the Einstein
radius relative to the effective radius. We will explore this issue
further in future papers of this series with a larger sample and
better data.

Finally, we also note that, despite the likely greater influence of
the environment relative to SLACS, the SL2S lenses do not require as
large a quadrupolar term as CLASS radio source or quasars
\citep{KKS97}.  This might be due to the fact that extended sources
are not as sensitive to the magnification bias that would boost the
fraction of highly magnified quads. In turn, the only unambiguous quad
systems here are J141137+565119 and J021737$-$051329. Two out of 16
lenses (excluding J221606$-$175131) are quads.
 
\section{Summary}\label{sect:conclude}

In this paper we presented gravitational lens models of a pilot sample
of 16 galaxy-scale lenses identified as part of the SL2S
Survey. {\em HST} imaging is available for 10/16 systems,
while for the others the modeling is based on ground-based CFHTLS
imaging. After removing the light from the foreground deflector we
used the surface brightness distribution of the background source to
constrain mass models of the deflector galaxy. For each system we
derived Einstein radius, position angle, and ellipticity of the mass
distribution described as a SIE. These
parameters are used in Paper II in combination with spectroscopic
information to study the relative distribution of stellar and dark
matter in ETGs and its evolution with cosmic time. In
this paper we focused on the relative orientation and ellipticity of
the luminous and total mass distribution, which does not require
spectroscopic information. We found that the ellipticity of mass
and light are correlated except for three outliers. In one case the
presence of a disk makes the light distribution significantly flatter
than the overall mass distribution. In two cases, the presence of a
nearby galaxy introduces significant external shear in the overall
mass distribution. In addition, we found that the position angle of
mass and light is on average aligned, albeit with rms scatter of
18-25 deg, significantly larger than what is found for the lower
redshift SLAC sample. We interpreted this scatter as most likely due
to substantial external shear, on average $\langle
\gamma_{\rm ext}\rangle = 0.12 \pm 0.05$, resulting from the environment
(physically related to the main deflector galaxy or along the line of
sight). The physical scale of the SL2S Einstein radii is greater that
for SLACS: $R_{\rm Ein}/R_{\rm eff}\simeq 1.1$ versus 0.5 for SLACS
\citep{Aug++10}. In addition, we found only two quad lenses in
this pilot sample, suggesting that magnification bias is less
effective in boosting the statistics of extended lensed sources as
compared with samples comprising point sources.

In Paper II we combine our lensing information with stellar kinematics
to infer the cosmic evolution of the mass density profile of massive
galaxies. Given the sample size in these first two papers of the
series, the role of environment and external shear cannot be explored
further, neither in terms of astrophysical signal, nor in terms of
systematic uncertainty on the mass density slope. However, the
follow-up imaging and spectroscopy of SL2S candidates are still ongoing
and we plan to investigate these issues with a much larger sample in
future papers of this series.  The parallel effort of measuring the
redshift of sources with Keck and XShooter on VLT will also allow us
to explore the population of lensed sources with greater detail.


\acknowledgments
We thank the anonymous referee for valuable suggestions and
a constructive report.
R.G. and F.B. were supported by the Centre National des Etudes Spatiales.
P.J.M. was given support by the TABASGO foundation in the form of a
research fellowship.
T.T. acknowledges support from the NSF through CAREER award NSF-0642621,
and from the Packard Foundation through a Packard Fellowship.
Based on observations obtained with MegaPrime/MegaCam, a joint project
of CFHT and CEA/DAPNIA, at the Canada--France--Hawaii Telescope (CFHT)
which is operated by the National Research Council (NRC) of Canada,
the Institut National des Sciences de l'Univers of the Centre National
de la Recherche Scientifique (CNRS) of France, and the University of
Hawaii.  This work is based in part on data products produced at
TERAPIX and the Canadian Astronomy Data Centre as part of the
Canada--France--Hawaii Telescope Legacy Survey, a collaborative project
of NRC and CNRS.
This research is supported by NASA through {\em Hubble Space Telescope}
programs GO-10876, GO-11289, GO-11588, and in part by the National
Science Foundation under grant No. PHY99-07949, and is based on
observations made with the NASA/ESA {\em Hubble Space Telescope} and
obtained at the Space Telescope Science Institute, which is operated
by the Association of Universities for Research in Astronomy, Inc.,
under NASA contract NAS 5-26555, and at the W.M. Keck Observatory,
which is operated as a scientific partnership among the California
Institute of Technology, the University of California, and the National
Aeronautics and Space Administration. The Observatory was made
possible by the generous financial support of the W.M. Keck
Foundation. The authors wish to recognize and acknowledge the very
significant cultural role and reverence that the summit of Mauna Kea
has always had within the indigenous Hawaiian community. We are most
fortunate to have the opportunity to conduct observations from this
mountain.


\bibliographystyle{aa}


\clearpage

\begin{landscape}
\renewcommand{\arraystretch}{1.50} 
\begin{deluxetable}{ccccccccccc}
\tabletypesize{\small}
\tablecaption{Summary of key observables.\label{table:main}}
\tabletypesize{\footnotesize}
\tablehead{
name & $z_{\rm d}$ &  $m_i$ & input &$R_{\rm Eff}$ & $R_{\rm Ein}$ & $q_{\rm SIE}$  & P.A.$_{\rm SIE}$  & $q_{\rm *}$  & P.A.$_{\rm *}$  & source mag \\
     &       &   (AB)      &  & (arcsec) & (arcsec)    & & & &  & 
}
\startdata
SL2SJ021411$-$040502  &  0.609 &  18.78  &   ACS/F606W &  $  0.94 \pm   0.05 $ & $  0.92 \pm   0.07 $ & $  0.33 \pm   0.08 $  &  $ 42.3 \pm   1.3$ & $  0.89 \pm   0.02 $  &  $ 53.6 \pm   4.0$  &  $24.30 \pm  0.12$  \\ 
SL2SJ021737$-$051329  &  0.646 &  19.47  &   ACS/F606W &  $  0.77 \pm   0.04 $ & $  1.27 \pm   0.06 $ & $  0.91 \pm   0.07 $  &  $124.3 \pm   2.4$ & $  0.92 \pm   0.02 $  &  $ 91.2 \pm   4.5$  &  $24.45 \pm  0.10$  \\ 
SL2SJ021902$-$082934  &  0.390 &  19.33  &   Megacam/g &  $  0.90 \pm   0.08 $ & $  0.92 \pm   0.06 $ & $  0.53 \pm   0.09 $  &  $102.0 \pm   4.1$ & $  0.73 \pm   0.01 $  &  $ 79.7 \pm   1.0$  &  $25.35 \pm  0.19$  \\ 
SL2SJ022056$-$063934  &  0.330 &  18.35  &   Megacam/g &  $  1.47 \pm   0.04 $ & $  1.25 \pm   0.06 $ & $  0.73 \pm   0.07 $  &  $ 76.6 \pm   2.1$ & $  0.57 \pm   0.01 $  &  $105.0 \pm   0.5$  &  $25.74 \pm  0.09$  \\ 
SL2SJ022511$-$045433  &  0.238 &  16.84  & WFPC2/F606W &  $  1.90 \pm   0.09 $ & $  1.77 \pm   0.09 $ & $  0.58 \pm   0.07 $  &  $ 26.3 \pm   0.2$ & $  0.75 \pm   0.01 $  &  $ 26.0 \pm   0.7$  &  $24.14 \pm  0.04$  \\ 
SL2SJ022610$-$042011  &  0.494 &  18.78  & WFPC2/F606W &  $  0.56 \pm   0.03 $ & $  1.15 \pm   0.06 $ & $  0.92 \pm   0.08 $  &  $  5.3 \pm  10.9$ & $  0.83 \pm   0.01 $  &  $ 56.2 \pm   1.2$  &  $25.10 \pm  0.08$  \\ 
SL2SJ022648$-$040610  &  0.766 &  20.01  & WFPC2/F606W &  $  1.20 \pm   0.06 $ & $  1.31 \pm   0.07 $ & $  0.82 \pm   0.07 $  &  $112.1 \pm   5.2$ & $  0.34 \pm   0.01 $  &  $131.8 \pm   0.6$  &  $26.17 \pm  0.09$  \\ 
SL2SJ022648$-$090421  &  0.456 &  18.32  &   Megacam/g &  $  1.30 \pm   0.04 $ & $  1.58 \pm   0.09 $ & $  0.83 \pm   0.08 $  &  $ 68.1 \pm   4.2$ & $  0.81 \pm   0.01 $  &  $ 61.6 \pm   0.9$  &  $27.13 \pm  0.14$  \\ 
SL2SJ023251$-$040823  &  0.352 &  18.58  & WFPC2/F606W &  $  0.81 \pm   0.04 $ & $  1.10 \pm   0.06 $ & $  0.80 \pm   0.07 $  &  $122.0 \pm   0.8$ & $  0.68 \pm   0.01 $  &  $110.2 \pm   0.8$  &  $26.39 \pm  0.08$  \\ 
SL2SJ140123$+$555705  &  0.526 &  19.14  & WFPC2/F606W &  $  0.76 \pm   0.04 $ & $  1.19 \pm   0.07 $ & $  0.49 \pm   0.08 $  &  $126.7 \pm   6.5$ & $  0.73 \pm   0.01 $  &  $136.0 \pm   1.1$  &  $25.19 \pm  0.17$  \\ 
SL2SJ140614$+$520252  &  0.480 &  18.29  &   Megacam/g &  $  2.15 \pm   0.16 $ & $  3.02 \pm   0.15 $ & $  0.29 \pm   0.07 $  &  $107.2 \pm   0.1$ & $  0.48 \pm   0.01 $  &  $120.4 \pm   0.6$  &  $23.50 \pm  0.03$  \\ 
SL2SJ141137$+$565119  &  0.322 &  18.49  &  WFC3/F475X &  $  0.76 \pm   0.04 $ & $  0.92 \pm   0.05 $ & $  0.90 \pm   0.07 $  &  $ 61.4 \pm   0.3$ & $  0.88 \pm   0.01 $  &  $ 17.7 \pm   1.4$  &  $25.86 \pm  0.04$  \\ 
SL2SJ220629$+$005728  &  0.704 &  20.59  &   Megacam/g &  $  2.25 \pm   0.24 $ & $  1.34 \pm   0.07 $ & $  0.46 \pm   0.07 $  &  $143.3 \pm   0.2$ & $  0.99 \pm   0.02 $  &  $108.8 \pm   5.1$  &  $25.81 \pm  0.08$  \\ 
SL2SJ221326$-$000946  &  0.338 &  20.00  & WFPC2/F606W &  $  0.41 \pm   0.02 $ & $  1.08 \pm   0.05 $ & $  0.19 \pm   0.07 $  &  $138.0 \pm   0.5$ & $  0.28 \pm   0.01 $  &  $148.4 \pm   0.6$  &  $25.14 \pm  0.04$  \\ 
SL2SJ221407$-$180712  &  0.650 &  20.43  & WFPC2/F606W &  $  0.57 \pm   0.03 $ & $  0.41 \pm   0.23 $ & $  0.69 \pm   0.16 $  &  $ 62.3 \pm  46.9$ & $  0.73 \pm   0.01 $  &  $ 57.5 \pm   0.8$  &  $25.29 \pm  0.32$  \\ 
SL2SJ221606$-$175131  &  0.860 &  20.68 &    Megacam/g &  $  0.93 \pm   0.08 $ &--&--&--& $ 0.87 \pm 0.01 $ & $ 57.4 \pm 3.1$ & $25.67 \pm 0.07$ \\
SL2SJ221929$-$001743  &  0.289 &  17.78  &   Megacam/g &  $  1.00 \pm   0.03 $ & $  0.74 \pm   0.14 $ & $  0.75 \pm   0.16 $  &  $ 85.3 \pm  35.5$ & $  0.74 \pm   0.01 $  &  $ 84.3 \pm   0.6$  &  $24.08 \pm  0.21$  \\ 
\enddata
\tablecomments{Effective radii and associated errors are estimated from the mean and standard deviation of a {\tt galfit} de Vaucouleurs fit in each of the $r$, $i$, and $z$ Megacam bands when HST imaging is not available. When one or more HST bands are available, $R_{\rm eff}$ is measured on the reddest one (See table~\ref{table:hst}) but the error is still obtained from the Megacam bands. (a) As discussed in Section 4.1, we could not model SL2SJ221606$-$175131 as a lens and therefore do not report lensing parameters for this system.}
\end{deluxetable}
\clearpage 
\end{landscape}

\end{document}